\documentclass[aps,prl,preprint,showpacs,superscriptaddress,showkeys]{revtex4-1}

\usepackage{amssymb}
\usepackage{amsmath}
\usepackage{amsfonts}
\usepackage{graphicx}
\usepackage{color}
\usepackage{xspace}
\usepackage{ulem}
\usepackage{mathtools}
\usepackage{hhline}

\newcommand{\AddrVandy}{Department of Physics and Astronomy, Vanderbilt University, Nashville, TN 37235, USA}

\begin{document}

{}

\title{Photon directional profile from stimulated decay of axion clouds with arbitrary momentum distributions}

\author{Liang Chen} \email{liang.chen@vanderbilt.edu}\affiliation{\AddrVandy}
\author{Thomas W. Kephart}    \email{thomas.w.kephart@vanderbilt.edu}\affiliation{\AddrVandy}

\date{\today}

\begin{abstract}
We model clusters of axions with spherically symmetric spacial but arbitrary momentum distributions and study the directional profile of photos produced in their evolution through spontaneous and stimulated decay of axions  via the process $a \rightarrow \gamma  \gamma$. Several specific examples are presented.
\end{abstract}

%\pacs{} 

\maketitle

\section{Introduction} 

The strong interactions conserve $CP$ invariance to a high degree of accuracy, but the Lagrangian for QCD 
$${\cal L} = \frac{1}{4}F^2 + \theta F{\tilde F} + m_F {\bar \psi} \psi$$
appears to have two allowed sources of $CP$ violation. 
One is the instant term $\theta F{\tilde F}$, and the other is the fermion mass matrix $m_F$. The full contribution can be written ${\bar \theta}= \theta+ arg[{det(m_F)}]$ and compared with the best experimental limit from the neutron electric dipole moment \cite{PDG} is ${\bar \theta} \le 10^{-10}$ radians.
To avoid such a fine-tuning, Peccei and Quinn \cite{Peccei:1977hh,Peccei:1977ur} proposed a solution to this so called strong $CP$ problem where ${\bar \theta}$ is promoted to a field $a$, the axions, who's potential when minimized relaxes the field (and affectively ${\bar \theta}$) to zero.
While this axion has a KeV mass   \cite{Weinberg:1977ma,Wilczek:1977pj} and interacts too strongly to be allowed by experiment, other ``invisible axion'' variants are lighter and allowed \cite{Kim:1979if}.

Invisible axions are weakly coupled light pseudo scalars that are a perfect dark matter candidate. Current experiments limit the  axion to the mass range $10^{-3}$ to $10^{-5}$ eV. They are  nonrelativistic if produced at the electroweak (EW) phase transition, and are unlike other particles in this mass range like neutrinos which were produced and thermalized in the Big Bang. The lower limit of the axion mass range is due to the fact that if they were lighter they would be so abundant that they would over close the Universe. Details can be found in   \cite{Abbott:1982af,Preskill:1982cy,Dine:1982ah}.
Early reviews of axions and the strong $CP$ problem and axions in cosmology and astrophysics are   \cite{Kim:1986ax,Cheng:1987gp,Raffelt:1990yz,Kolb:1990vq}.

The physics of the axion is similar to the $\pi^0$ in that they are both pseudo scalars and can decay to two photons through the one loop triangle anomaly diagram related to breaking  the axial $U(1)$ global symmetry of massless QCD. The coupling of the axion field $a(x)$ to photons is 
$${\cal L}_{a \rightarrow \gamma \gamma} =\frac{\alpha K}{8\pi F_a} a(x)F{\tilde F} $$
where $\alpha$ is the fine structure constant, $K$ is an $O(1)$ model dependent constant and $F_a$ is the axion decay constant which can be related to the pion mass and decay constant through $m_a F_a \approx m_{\pi} F_{\pi}$ where $m_a$ is the axion mass. The axion lifetime scales with the $\pi^0$ lifetime
$$\tau_a \approx \left(\frac{m_{\pi}}{m_a}\right)^5 \tau_{\pi^0} \sim 10^8 \left(\frac{eV}{m_a}\right)^5 Gyr$$
so axions are very long lived compared to the age of the Universe.
Since axions are long lived and copiously produced at rest in the EW phase transition, they are a candidate for cold dark matter and can be the source of early universe density perturbation that grow in a way consistent with observations of the cosmic microwave background (CMB), which is unlike the results for free streaming hot dark matter perturbation that do not grow at small length scales.

Axions in vacuum can undergo spontaneous decay $a \rightarrow \gamma \gamma$, and if their density is high enough, stimulated decay can cause lasing to commence and rapidly deplete the axion number density, while the photon number grows exponentially. Sufficiently dense regions of axions can release enough monochromatic electromagnetic energy for possible detection \cite{Kephart:1986vc,Tkachev:1987cd}. The case of lasing  axions for  a spherical symmetric distribution, both in momentum and coordinate space has been considered in  \cite{Kephart:1986vc,Kephart:1994uy,Tkachev:1987cd}. However, resent studies of axions as dark matter suggests they may be in several other phase space configurations. E.g., they may fall along caustics in galaxies   \cite{Sikivie:1997ng,Duffy:2008dk}, or they may be produced by superradiance, a Penrose type of process, around black holes \cite{Rosa:2017ury} and fill a hydrogen-like orbit with quantum numbers $(n,l,m) = (2,1,1).$
(Since axions are bosons there is no fundamental limit on their numbers in this state.)
For these reason we have decided to begin an exploration of the general case of lasing for arbitrary axion distributions in terms of a spherical harmonic expansion.

The axion mass range can be expanded somewhat by fine tuning, but this begins to defeat the purpose of introducing the axions in the first place to avoid fine tuning ${\bar\theta}$. The range can also be loosened by considering axion-like particles not necessarily designed to solve the strong $CP$ problem. The analysis we present here can be applied to any of these cases as long as the axion or axion-like particles have a dominant decay mode to two photons.

In this paper we focus on arbitrary momentum distributions while keeping the coordinate space distribution spherically symmetric. In particular we write a general expansion of momentum modes in spherical harmonics. Many of the algebraic details are relagated to the appendices.
Once we have derived the general form we then consider a few specific examples of physical interest. In future work we plan to also relax the coordinate space spherical symmetry requirement.

%%%%%%%%%%%%%%%%%%%%%%%%%%%%%%%%%%%%%%%%%%%%%%%%%%%%%%%%%%%%%%%%%%%%%%%%%%%%%%%%%%%%%%%%

\section{Setup and preparation}

To fix notation and set the stage for our analysis we first
 review the setup of the sphericrally symmetric phase space axion cluster model  \cite{Kephart:1986vc,Kephart:1994uy}. Axions of mass density $\rho_a$ are produced nonthermally during the QCD phase transition with a cosmological
density parameter today \cite{Abbott:1982af,Preskill:1982cy,Dine:1982ah} of
$$\Omega_a\equiv \frac{\rho_a}{\rho_c} \cong \left(\frac{10^{-5}\,\, eV}{m_a} \right)^{7/6},$$
where $\rho_c$ is the critical density.
Hence axions are a dominant dark matter  (DM) candidate   if $m_a \sim 10^{-5}$ eV.

Our purpose here is to give detailed calculations of the
stimulated emission rate for axion clusters. There
are many possible sources of initial photons of frequency in the $ \sim 10^{-5}$ eV range,
therefore spontaneous decays are not required to start 
the lasing process, although spontaneous decay of axions 
is a lower bound on axion cluster luminosity. 
We neglect all normal matter in the cluster, hence we  neglect the 
attenuation of photons by nonaxionic matter, as well as any other 
affects due to the local environment.

For any species of particles with occupation number
 $f(p, r, t)$, the particle number density is
$$n(r, t)= \int \frac{d^3p}{8\pi^3}f(p, r, t),$$
and the total particle number of these particles in volume
$V$ is
$$N = \int_V d^3r\,\, n(r, t).$$

Angular momentum conservation requires that a pair of photons 
emitted by decay of a spin zero particle
(scalar or pseudoscalar) have the same helicity.
The rate of change
of the photon number density of helicity $\lambda=\pm 1$
within an axion cluster, due to the process $a\leftrightarrow \gamma \gamma$ can be written as a Lorentz invariant phase space integral
$$\frac{dn_{\lambda}}{dt}=\int dX^{(3)}_{LIPS}[f_a(1+f_{1\lambda})(1+f_{2\lambda})-f_{1\lambda}f_{2\lambda}(1+f_a)]|M(a\rightarrow \gamma \gamma)|^2$$
where $f_a=f_a({\vec p})$, $f_{i\lambda}=f_{i\lambda}({\vec k}_i)$ and $M=M(a\rightarrow\gamma(+)\gamma(+))=M(a\rightarrow\gamma(-)\gamma(-))$. In detail, the phase space integral is
$$ \int dX^{(3)}_{LIPS} = \int \frac{d^3p}{(2\pi)^3 2p^0}  \int \frac{d^3k_1}{(2\pi)^3 2k_1^0}   \int \frac{d^3k_2}{(2\pi)^3 2k_2^0} \delta^{4}(p-k_1-k_2),$$
where $p$ is the axion momentum and $k_1$ and $k_2$ are the momenta of the photons.

 Upon defining $k=k^0_2$ and using the above results, we arrive at the rate equation for the photon occupation number 
 $$2k\frac{df_{\lambda}({\vec k})}{dt}=\frac{4m_a\Gamma_a}{\pi}\int  \frac{d^3k_1}{ 2k_1^0} \frac{d^3p}{ 2p^0}\delta^{4}(p-k_1-k)f_a({\vec p})[1+f_{\lambda}({\vec k})+f_{\lambda}({\vec k}_1)]-f_{\lambda}({\vec k})f_{\lambda}({\vec k}_1)$$

In the case of spherical symmetry we set $f({\vec q})=f(|{\vec q}|)=f(q)$ and the intergral simplifies to give

 $$\frac{df_{\lambda}(k)}{dt}=\frac{4m_a\Gamma_a}{\pi}\int^\infty_{k+\frac{m_a^2}{4k}}  dp^0\delta^{4}(p-k_1-k)
 f_a(p^0)[1+f_{\lambda}(k)+f_{p^0-k}]-f_{\lambda}(k)f_{\lambda}(p^0-k)$$

In \cite{Kephart:1986vc,Kephart:1994uy}   a specific model with spherical symmetry was choosen  where  the initial axions were contained in a ball of radius $R$, with a maximum momentum value of $p_{\textrm{\tiny max}} \approx m_a\beta$ in the nonrelativistic case. I.e., the initial axion occupation number was chosen to be
 $$f_a(p,r)=C_a\Theta(p_{max}-p)\Theta(R-r)$$
 Where $C_a$ is a constant that can be written in terms of axion mass and initial number density. Details of this model were worked out in
  \cite{Kephart:1986vc,Kephart:1994uy} and will be recovered below
when we take the spherical symmetric limit of the general case. \\

 The direction of axion momentum can be characterized by the infintestimal solid momentum angle $\Omega_p=(\theta_p,\phi_p)$. $f_a(p, r, \Omega_p, t)$ and $n_a(r, \Omega_p, t)$ are the axion occupation number and axion number density, respectively. A integral of $f_a(p, r, \Omega_p, t)$  over $p$ or $(p, \Omega_p)$ gives number density $n_a(r, \Omega_p, t)$ or $n_a(r, t)$.\\

The photons produced in  the decay of axions are initially contained in the ball of radius $R$, and in a momentum spherical shell of inner and outer radius $k_-  =\frac{m_a\gamma}{2}(1-\beta)$ and $k_+ = \frac{m_a\gamma}{2}(1+\beta) $, respectively. $f_\lambda(k, r, \theta, t)$ and $n_\lambda( r, \theta, t)$ are the photon occupation number and photon number density, respectively, of helicity $\lambda=\pm1$. We assume that the number density of each helicity state is the same, so the total photon number density $n_\gamma$ can be written as $ n_\gamma( r, \theta, t) = n_+( r, \theta, t) + n_-( r, \theta, t)=2n_{\lambda}( r, \theta, t)$.

Since the momentum distribution of axions is assumed to be spherically symmetric,  there is an equal chance for an axion to be moving in any  direction. Likewise the lasing process is equally likely to commence in any direction.
In  \cite{Kephart:1986vc,Kephart:1994uy} this model was used to give lasing bounds and find the stimulated decay rate equations for axions and photons. Numerical results were also given for the evolution of such an axion cluster.

Here we want to investigate how the momenta of photons are distributed if a general non-spherically symmetric distribution of the axion momenta  is specified. 
We do this by generalizing the  model of  \cite{Kephart:1986vc,Kephart:1994uy} and proceeding without making this assumption of spherical symmetry in momentum space. We proceed by expanding in spherical harmonics and then  integrating over all  angles of photon momentum.

\section{Evolution equations for individual components of photon occupation number}

Once we relax spherical symmetry we must take angular dependencies into account. There are several sets of angles to consider, those for both axion and photon position and momentum, hence a potential total of eight angles on which our results may depend. Let us begin with the axions.  

No spherical symmetry in momentum space means  $f_a( \vec{p} ) = f_a ( p, \theta_p, \phi_p )$ is a general function of angles $\theta_p$ and $ \phi_p$. Since $f_a( \vec{p} )$ is the occupation number of axions, it has to be integrable. 
\subsection{Harmonic Expansion}
We can write $f_a( \vec{p} )$ as the square of a square integrable function formed in a complex spherical harmonics expansion,
\begin{flalign}\nonumber
f_a( \vec{p} ) &= [\sum_{l'm'}a_{l'}^{m'}(p,t)Y_{l'}^{m'}(\Omega_p)]^*[\sum_{lm}a_l^m(p,t)Y_l^m(\Omega_p)] \\ \nonumber
%&= \sum_{l'lm'm}a_{l'}^{m'*}a_l^m(-1)^{m'}Y_{l'}^{-m'}(\Omega_p)Y_l^m(\Omega_p)   \\ \nonumber
%&= \sum_{l'lm'm}(-1)^{m'}a_{l'}^{-m'*}a_l^mY_{l'}^{m'}(\Omega_p)Y_l^m(\Omega_p)   \\ \nonumber
&= \sum_{l'lm'm}\sum_L (-1)^{m'}a_{l'}^{-m'*}a_l^m \sqrt{\frac{(2l'+1)(2l+1)}{4\pi(2L+1)}}  \times  \\ \nonumber
& C(l',l,L|0,0,0)C(l',l,L|m',m,m'+m)Y_L^{m'+m}(\Omega_p).
\end{flalign}
where the $Y_l^m$s are complex spherical harmonics, and the $C(l,l',l''|m,m'm'')$s are Clebsch-Gorden coefficients.

By regrouping and renaming coefficients, $f_a( \vec{p} )$ can be written directly as a real spherical harmonic expansion,
\begin{flalign}  \label{f_a}
f_a( \vec{p} ) &= \sum_{lm}a_{lm}(p,t)Y_{lm}(\Omega_p) ~.
\end{flalign}
Similarly, photon occupation numbers can also be written as
\begin{flalign} 
f_\lambda( \vec{k} ) &= \sum_{lm}b_{lm}(k,t)Y_{lm}(\Omega_k) ~, \\  
f_\lambda( \vec{k}_1 ) &= \sum_{lm}b_{lm}(k_1,t)Y_{lm}(\Omega_{k_1})  ~.
\end{flalign}

The following calculations do not put any restrictions on the coefficients $a_{lm}$'s or $b_{lm}$'s, neither do the calculations require these coefficients to be positive or negative. But in real physical world, only positive occupation numbers are allowed. In short, these coefficients can be of any value, but we should scrutinize the resulting occupation numbers so that they are positive and have real world meaning. For example, occupation number $f_a( \vec{p} ) = a_{10}(p,t)Y_{10}(\Omega_p)$ can not describe any real scenario since it is negative in half of phase space, but occupation number $f_a( \vec{p} ) = a_{00}(p,t)Y_{00}(\Omega_p) + a_{10}(p,t)Y_{10}(\Omega_p)$ may be allowed since it can be positive everywhere by adjusting the coefficients $a_{00}$ and $a_{10}$. Nevertheless the following calculations are applicable to both of these occupation numbers, whether they have real physical meaning or not.

\subsection{Integration over $k_1, \Omega_{k_1}$}
We start our analysis from the general  rate equation given in \cite{Kephart:1994uy}, which describes the evolution process of spontaneous and stimulated decay of axions to photons, and the back reaction of photons to axions.
\begin{flalign} \label{10Kephart}
 2k \frac{d f_\lambda ( \vec{k} )  }{dt}  =&  \frac{4 m_{a} \Gamma_{a} }{ \pi }  \int \frac{d^3  k_1   }{ 2k_1^0 } \frac{d^3  p   }{ 2p^0 }  \delta^4 (p-k-k_1) \times \\  \nonumber
   &  \{  f_a  ( \vec{p} ) [ 1+ f_\lambda ( \vec{k} ) + f_\lambda ( \vec{k}_1 ) ] - f_\lambda ( \vec{k} )f_\lambda ( \vec{k}_1 ) \}  ~.
\end{flalign}
where $\Gamma_a= 1/\tau_a$ is the axion decay width. Writing out the differential $d^3  k_1$ and the $\delta$-function explicitly, we have
\begin{flalign}  \nonumber  
2k \frac{d f_\lambda ( \vec{k} )  }{dt} =&  \frac{ m_{a} \Gamma_{a} }{ \pi }  
\int \frac{(k_1)^2 dk_1 \sin\theta_{k_1}  d\theta_{k_1} d\phi_{k_1}   }{ k_1^0 }  \frac{d^3  p   }{ p^0 }   \\ \nonumber
&     \frac{1}{(k_1)^2 \sin\theta_{k_1}}   \delta (p^0-k^0-k_1^0)   \delta (|\vec{p}-\vec{k}|-k_1) \\ \nonumber
&     \times     \delta (\theta_{\vec{p}-\vec{k}}-\theta_{\vec{k}_1})     \delta (\phi_{\vec{p}-\vec{k}}-\phi_{\vec{k}_1})      \\ \nonumber
&     \times \{  f_a  ( \vec{p} ) [ 1+ f_\lambda (\vec{k} ) + f_\lambda (\vec{k}_1 ) ] - f_\lambda ( \vec{k} )f_\lambda ( \vec{k}_1 ) \} ~,
\end{flalign}
canceling  common factors and substituting from the harmonic expansions, we find
\begin{flalign}  \nonumber  
& 2k \frac{d f_\lambda ( \vec{k} )  }{dt}   
 =\frac{ m_{a} \Gamma_{a} }{ \pi }
\int \frac{ dk_1     }{ k_1^0 } \frac{d^3  p   }{ p^0 }  \delta (p^0-k^0-k_1^0)  \delta (|\vec{p}-\vec{k}|-k_1)   \\ \nonumber
&     \times \delta (\theta_{\vec{p}-\vec{k}}-\theta_{\vec{k}_1})  \delta (\phi_{\vec{p}-\vec{k}}-\phi_{\vec{k}_1})    
      \{  f_a  ( \vec{p} ) [ 1+ f_\lambda (\vec{k} ) + \sum_{lm}b_{lm}(k_1,t)Y_{lm}(\Omega_{k_1}) ]   \\ \nonumber
&     - f_\lambda ( \vec{k} ) \sum_{lm}b_{lm}(k_1,t)Y_{lm}(\Omega_{k_1})   \}   d\theta_{k_1} d\phi_{k_1}  ~.
\end{flalign}
We can do the integration over $\theta_{k_1}$ and $\phi_{k_1}$ most efficiently by changing from $\Omega_{k_1}$ to $\Omega_{\vec{p}-\vec{k}}$.
Upon use the fact that  for photons $ k_1^0 = k_1 $ and $ k^0 = k $ along with the identity
\begin{flalign}  \nonumber  
\int \delta(x-y) \delta(x-z) dx = \delta(y-z)   ~,
\end{flalign}
we find
\begin{flalign}  \nonumber  
& 2k \frac{d f_\lambda ( \vec{k} )  }{dt}   
=  \frac{ m_{a} \Gamma_{a} }{ \pi }
\int  \frac{d^3  p   }{ p^0 (p^0-k) }    \delta [|\vec{p}-\vec{k}|- (p^0-k) ]  \\ \nonumber
&     \times \{  f_a  ( \vec{p} ) [ 1+ f_\lambda (\vec{k} ) + \sum_{lm}b_{lm}(p^0-k ,t)Y_{lm}(\Omega_{\vec{p}-\vec{k}}) ]   \\ \label{10Kephart-1}
&    - f_\lambda ( \vec{k} ) \sum_{lm}b_{lm}(p^0-k,t)Y_{lm}(\Omega_{\vec{p}-\vec{k}})   \}     ~.
\end{flalign}
\subsection{Integration over $\phi_p$ : preparation }

Substituting $d^3p=p^2dp\, d\Omega_p = \sqrt{(p^0)^2-m_a^2} p^0 d p^0 \, d\Omega_p$ and various relations from   Appendix A into  equation \eqref{10Kephart-1}, we find
\begin{flalign}  \nonumber  
&  2k \frac{d f_\lambda ( \vec{k} )  }{dt}    
= \frac{ m_{a} \Gamma_{a} }{ \pi } 
     \int  \frac{ \sqrt{(p^0)^2-m_a^2} p^0 d p^0 \, d\Omega_p    }{ p^0 (p^0-k) }  \\ \nonumber
&     \times \frac{p^0-k}{k\sqrt{(p^0)^2-m_a^2}  \sin\theta_{p0}}   \delta( \theta_p - \theta_{p0}  ) \\ \nonumber
&     \times \{  f_a  ( \vec{p} ) [ 1+ f_\lambda (\vec{k} ) + \sum_{lm}b_{lm}(p^0-k ,t)Y_{lm}(\Omega_{\vec{p}-\vec{k}}) ]   \\ \nonumber
&       - f_\lambda ( \vec{k} ) \sum_{lm}b_{lm}(p^0-k,t)Y_{lm}(\Omega_{\vec{p}-\vec{k}})   \}     ~.
\end{flalign}

Canceling common factors and writing the equation in a form that is suitable for doing the two angular integrations, we have
\begin{flalign}  \nonumber  
&  2k \frac{d f_\lambda ( \vec{k} )  }{dt}    
=  \frac{ m_{a} \Gamma_{a} }{ \pi }
     \int  \frac{   d p^0 \,  \delta( \theta_p - \theta_{p0}  )    }{k  \sin\theta_{p0}}   \sin\theta_p d\theta_p   \\ \nonumber
&     \times \{    [ 1+ f_\lambda (\vec{k} ) ]  \int  f_a  ( \vec{p} )  \, d\phi_p \\ \nonumber
&   + \int  f_a  ( \vec{p} )  \sum_{lm}b_{lm}(p^0-k ,t)Y_{lm}(\Omega_{\vec{p}-\vec{k}}) \,  d\phi_p   \\ \label{10Kephart-2}
&     -   f_\lambda ( \vec{k} )   \int   \sum_{lm}b_{lm}(p^0-k,t)Y_{lm}(\Omega_{\vec{p}-\vec{k}})   \,   d\phi_p  \}     ~.
\end{flalign}

The momentum space angular integration requires some algebra which can be found in Appendix C.
Adopting $\theta_{p1}$ notation from eq. eq. \eqref{p1} in Appendix C, eq. \eqref{10Kephart-2} become
\begin{flalign}  \nonumber  
&  2k \frac{d f_\lambda ( \vec{k} )  }{dt}    \\ \nonumber 
=&   2m_{a} \Gamma_{a}    \int  \frac{   d p^0     }{k  }  ~     \{    [ 1+ f_\lambda (\vec{k} ) ]   
       \sum_{l}a_{l0}(p,t)   Y_{l0}( \theta_{p0} )  +  \\ \nonumber 
                &    \sum_{l'lm}a_{l'm}(p,t) b_{lm}(p^0-k ,t) Y_{l'}^{m}(\theta_{p0},0)  Y_{l}^{m}(\theta_{p1},0)   \\ \label{10Kephart-3}
                &     -   f_\lambda ( \vec{k} )  \times   \sum_{l}b_{l0}(p^0-k,t)  Y_{l0}(\theta_{p1})  \}  ~.
\end{flalign}
\subsection{Evolution of individual components  }
The $\delta$-function in \eqref{10Kephart} imposes conservation of 4-momentum, since the integration has been carried out, $k$ and $p^0$ in \eqref{10Kephart-3} of Appendix C are independent. Hence, the $k$ in the denominator of \eqref{10Kephart-3} can be taken outside of the integral,
\begin{flalign}  \nonumber  
&  \frac{d f_\lambda ( \vec{k} )  }{dt}  =   \sum_{lm} \frac{ d b_{lm}(k,t) }{ dt }  Y_{lm}(\Omega_k)     \\ \nonumber 
=&  \frac{  m_{a} \Gamma_{a} }{k^2  }
                      \int  d p^0      ~  \{       \sum_{l}a_{l0}(p,t)   Y_{l0}( \theta_{p0} )  + \\ \nonumber 
                &   \sum_{l'lm}a_{l'm}(p,t) b_{lm}(p^0-k ,t) Y_{l'}^{m}(\theta_{p0},0)  Y_{l}^{m}(\theta_{p1},0)  +  \\ \nonumber
                &      f_\lambda ( \vec{k} )    [ \sum_{l}a_{l0}(p,t)   Y_{l0}( \theta_{p0} )   
                               -\sum_{l}b_{l0}(p^0-k,t)  Y_{l0}(\theta_{p1})  ] \}  ~.
\end{flalign}
Upon writing out the $\Omega_k$ dependence of RHS of the equation above explicitly, and identifying  the coefficients of $Y_{lm}(\Omega_k)$, we obtain differential an equations for components $b_{lm}(k,t)$,

\begin{flalign}  \nonumber  
&  \frac{ d b_{lm}(k,t) }{ dt } =   \\ \nonumber 
& \frac{  m_{a} \Gamma_{a} }{k^2  }   \int     d p^0 \{ \delta_{l0}\delta_{m0} 2\sqrt\pi   
        [  \sum_{l'}a_{l'0}(p,t)   Y_{l'0}( \theta_{p0} )  + \\ \nonumber 
                                                      &   \sum_{l'l''m'}a_{l'm'}(p,t) b_{l''m'}(p^0-k ,t) Y_{l'}^{m'}(\theta_{p0},0)  Y_{l''}^{m'}(\theta_{p1},0) ] +  \\ \nonumber
                                                      &   b_{lm}(k,t)     [ \sum_{l'}a_{l'0}(p,t)   Y_{l'0}( \theta_{p0} ) -\sum_{l'}b_{l'0}(p^0-k,t)  Y_{l'0}(\theta_{p1})  ] \}  .
\end{flalign}
Replacing $p^0$ with $k_1+k$, also using the following substitution,
\begin{flalign}  \nonumber  
\int_{k+\frac{m_a^2}{4k}} dp^0  \quad &\Leftrightarrow \quad   \int_{\frac{m_a^2}{4k}} dk_1   ~,
\end{flalign}
we find our  final form of the evolution equations for individual components of occupation number,
\begin{flalign} \label{occupation number components equations}
&\frac{ d b_{lm}(k,t) }{ dt } 
= \frac{  m_{a} \Gamma_{a} }{k^2  }       \int_{\frac{m_a^2}{4k}} dk_1   \{ \delta_{l0}\delta_{m0} 2\sqrt\pi 
                                                              [ \sum_{l'}  a_{l'0}( p,t)   Y_{l'0}( \theta_{p0} )      \\ \nonumber 
                   & + \sum_{l'l''m'}  a_{l'm'}(p,t) b_{l''m'}(k_1 ,t) Y_{l'}^{m'}(\theta_{p0},0)  Y_{l''}^{m'}(\theta_{p1},0) ]    \\ \nonumber
                    &+ b_{lm}(k,t)    [ \sum_{l'}a_{l'0}(p,t)   Y_{l'0}( \theta_{p0} ) -\sum_{l'}b_{l'0}(k_1,t)  Y_{l'0}(\theta_{p1})  ] \}   ~.
\end{flalign}
The terms inside curly bracket correspond to spontaneous decay, half of stimulated decay, the other half of stimulated decay, and photon annihilation, respectively.
Spontaneous decay and half of the stimulated decay of axion only contribute  to the order of $l=0,m=0$. 
In eq.\eqref{occupation number components equations}, we see that only axions of $m=0$ momentum configurations take part in spontaneous decay.
$l\neq0,\,m\neq0$ photon momentum distribution can only be produced from half of the stimulated decay and back reaction.
\section{Evolution equations for individual components of axion and photon number densities}
Axion and photon occupation numbers depend on position. This fact was not included in previous discussion, since it is not involved in the dynamical process. Here we will assume that spherical symmetry is  present in position space and axions sit inside a ball of radius $R$,
\begin{flalign}\nonumber
& f_a( \vec{p} ) = f_a( t, p, \Omega_p, \vec{r})  = \sum_{lm}  a_{lm}(t, p, \vec{r})Y_{lm}(\Omega_p)    ~,
\end{flalign}
where the occupation number components can be written 
\begin{flalign} \label{a} 
&a_{lm}(t, p, \vec{r}) = a_{lm}(t) \Theta(p_{\mbox{\tiny max}}-p)  \, \Theta(R-r)     ~.
\end{flalign}
Hence, the axion number density is
\begin{flalign} \nonumber
& n_a(t, \Omega_p, \vec{r}) = \sum_{lm}n^a_{lm}(t)  \, \Theta(R-r) Y_{lm}(\Omega_p)    ~,
\end{flalign}
where the  components $n^a_{lm}(t)$ and $a_{lm}(t)$ are related through
\begin{flalign} \nonumber
  n^a_{lm}(t) =&\int\frac{p^2dp}{(2\pi)^3}  a_{lm}(t) \Theta(p_{\mbox{\tiny max}}-p)     
=\frac{ (m_a\beta)^3 }{ 24\pi^3 }  a_{lm}(t)     ~,   \end{flalign}
or
\begin{flalign} \label{axion-an}
  a_{lm}(t)=\frac{24\pi^3 n^a_{lm}(t)}{ m_a^3 \beta^3}   ~.
\end{flalign}

Photo occupation number and number density can be treated in a similar fashion. We write
\begin{flalign}\nonumber
&f_\lambda( \vec{k} ) = f_\lambda( t, k, \Omega_k, \vec{r} ) = \sum_{lm}b_{lm}(t, k, \vec{r})Y_{lm}(\Omega_k)  ~, \end{flalign}
where
\begin{flalign}\label{b} 
b_{lm}(t, k, \vec{r}) = b_{lm}(t) \Theta(k_+-k) \Theta(k-k_-) \Theta(R-r)   ~, 
\end{flalign}
and
\begin{flalign}\nonumber
&n_\lambda(t, \vec{r}, \Omega_k)=   \sum_{lm}n^\lambda_{lm}(t)  \, \Theta(R-r) Y_{lm}(\Omega_k)  ~.
\end{flalign}
The components $n^\lambda_{lm}(t)$ and $b_{lm}(t)$ are related through
\begin{flalign}\nonumber
&  n^\lambda_{lm}(t) =\int\frac{k^2dk}{(2\pi)^3} b_{lm}(t) \Theta(k_+-k) \Theta(k-k_-)   ~,  
\end{flalign}
or
\begin{flalign}\label{photon-bn}
 b_{lm}(t)=\frac{32\pi^3 n^\lambda_{lm}(t)}{ m_a^3 \beta} ~.
\end{flalign}

\bigskip

In Appendix C, we replaced all the occupation number components in \eqref{occupation number components equations} with the corresponding number density components and integrated over the momentum space, which gives us the evolution equations for the individual components of number density of photon, normal axion, and sterile axion.
\begin{flalign}  \label{photon number density components equations}  
& \frac{d n^\gamma_{lm}(t) }{dt}        =    \frac{  2 \Gamma_{a} }{  m_a^2 \beta^2  }   \{ \delta_{l0}\delta_{m0} \sqrt\pi  [  \sum_{l'} \frac{6 }{ \beta}   K^{0}_{l'}   n^a_{l'0}(t)   \\ \nonumber 
                                                      &  + \frac{ 96\pi^3 }{ m_a^3 \beta^2}  \sum_{l'l''m'}   n^a_{l'm'}(t)  n^\gamma_{l''m'}(t)   K^{01}_{l'l''m'}   ]   \\ \nonumber
                                                      &  + \frac{ 16\pi^3 n^\gamma_{lm}(t)}{ m_a^3 }   \times   [ \frac{ 3 }{  \beta^2} \sum_{l'}  n^a_{l'0}(t)   K^{0}_{l'} - 2\sum_{l'} n^\gamma_{l'0}(t)   B^1_{l'}  ] \}      \\ \nonumber
                                                      &- \frac{3c}{2R}n^\gamma_{lm}(t)  ~.
\end{flalign}
\begin{flalign}  \label{normal axion number density components equations}  
& \frac{d n^a_{lm}(t) }{dt}        =   - \frac{   \Gamma_{a} }{  m_a^2 \beta^2  }   \{ \delta_{l0}\delta_{m0} \sqrt\pi  [  \sum_{l'} \frac{6 }{ \beta}   K^{0}_{l'}   n^a_{l'0}(t)   \\ \nonumber 
                                                      &  + \frac{ 96\pi^3 }{ m_a^3 \beta^2}  \sum_{l'l''m'}   n^a_{l'm'}(t)  n^\gamma_{l''m'}(t)   K^{01}_{l'l''m'}   ]   \\ \nonumber
                                                      &  + \frac{ 16\pi^3 n^\gamma_{lm}(t)}{ m_a^3 }   \times   [ \frac{ 3 }{  \beta^2} \sum_{l'}  n^a_{l'0}(t)   K^{0}_{l'} - 2\sum_{l'} n^\gamma_{l'0}(t)   N^1_{l'}  ] \}      ~.
\end{flalign}
\begin{flalign}  \label{sterile axion number density components equations}  
& \frac{d n^{as}_{lm}(t) }{dt}        =    \frac{   \Gamma_{a} }{  m_a^2 \beta^2  }   \frac{ 16\pi^3 n^\gamma_{lm}(t)}{ m_a^3 }   \times   2\sum_{l'} n^\gamma_{l'0}(t)   S^1_{l'}     ~.
\end{flalign}
The $K^{0}_{l'}$ are defined to be constant coefficients describing spontaneous and half of stimulated decay. $K^{01}_{l'l''m'}$ are constant coefficients describing the other half of stimulated decay. The $B^1_{l'}$ are constant coefficients describing  back reaction of photons, which when necessary, can be split into $N^1_{l'}$ and $S^1_{l'}$ the part of back reactions that produce normal axions and sterile axions respectively, and hence $B^1_{l'}=  N^1_{l'} + S^1_{l'}   $.

\section{Examples}
\subsection{$Y_{00}$ momentum distribution}
The choice $Y_{00}$ corresponds to the momentum distribution of axion with no preferred direction. The number density of axion for such a momentum distribution is
\begin{flalign}  \nonumber  
n_a(t, \vec{r}, \Omega_p) &=\sum_{lm}n^a_{lm}(t)  \, \Theta(R-r) Y_{lm}(\Omega_p)       \\ \nonumber
                                                      &=n^a_{00}(t) \, \Theta(R-r) Y_{00}(\Omega_p)                       ~.
\end{flalign}
Thus, for any $l\neq0$, $m\neq0$, the number density component is zero,
\begin{flalign}  \nonumber   n^a_{lm}(t)=0 \quad (lm\neq00) ~. \end{flalign}
This means that the evolution equations \eqref{normal axion number density components equations} for these components reduce to
\begin{flalign}  \nonumber  
 \frac{d n^a_{lm}(t) }{dt}        &=   - \frac{   \Gamma_{a} }{  m_a^2 \beta^2  }  \times \frac{ 16\pi^3 n^\gamma_{lm}(t)}{ m_a^3 }                                                                                                                                                                                                                                                                                \times  \\ \nonumber
                                                         &\qquad    [ \frac{ 3 }{  \beta^2} \sum_{l'}  n^a_{l'0}(t)   K^{0}_{l'} - 2\sum_{l'} n^\gamma_{l'0}(t)   N^1_{l'}  ]    = 0    ~.
\end{flalign}
There are two cases regarding the solution of the equation above.\\

The first case is
\begin{flalign}  \nonumber   n^\gamma_{lm}(t)=\delta_{l0}\delta_{m0}n^\gamma_{00}(t) ~, \end{flalign}
where all the photon number density components $n^\gamma_{lm}$ vanish except  $l=0$, and$m=0$,
\begin{flalign}  \nonumber  n_\gamma(t, \vec{r}, \Omega_k)=n^\gamma_{00}(t) \, \Theta(R-r) Y_{00}(\Omega_k)  ~. \end{flalign}
Since the photon field is fixed in the  $Y_{00}$ momentum distribution, the exact form of number density of normal axion, photon and sterile axion can be obtained by solving the evolution equations \eqref{photon number density components equations}, \eqref{normal axion number density components equations} and \eqref{sterile axion number density components equations}. In this first case we find
\begin{flalign}  \nonumber
 \frac{d n^\gamma_{00}(t) }{dt}        =&    \frac{  2 \Gamma_{a} }{  m_a^2 \beta^2  }   \{  \sqrt\pi  [   \frac{6 }{ \beta}   K^{0}_0   n^a_{00}(t)   
                                                        + \frac{ 96\pi^3 }{ m_a^3 \beta^2}     n^a_{00}(t)  n^\gamma_{00}(t)   K^{01}_{000}   ]   \\  \label{A}
                                                      &  + \frac{ 16\pi^3 n^\gamma_{00}(t)}{ m_a^3 }   \times   [ \frac{ 3 }{  \beta^2}   n^a_{00}(t)   K^{0}_0 - 2 n^\gamma_{00}(t)   B^1_0  ] \}       
                                                       - \frac{3c}{2R}n^\gamma_{00}(t) ~.                                                                                                      
 \\ \nonumber
 \frac{d n^a_{00}(t) }{dt}        =&   - \frac{   \Gamma_{a} }{  m_a^2 \beta^2  }   \{  \sqrt\pi  [   \frac{6 }{ \beta}   K^{0}_0   n^a_{00}(t)   
                                                         + \frac{ 96\pi^3 }{ m_a^3 \beta^2}  \  n^a_{00}(t)  n^\gamma_{00}(t)   K^{01}_{000}   ]   \\     \label{B}    
                                                      &  + \frac{ 16\pi^3 n^\gamma_{00}(t)}{ m_a^3 }   \times   [ \frac{ 3 }{  \beta^2}   n^a_{00}(t)   K^{0}_0 - 2 n^\gamma_{00}(t)   N^1_0  ] \}       ~.
 \\  \label{C}
 \frac{d n^{as}_{00}(t) }{dt}        =&    \frac{   \Gamma_{a} }{  m_a^2 \beta^2  }   \frac{ 16\pi^3 n^\gamma_{00}(t)}{ m_a^3 }   \times   2  n^\gamma_{00}(t)   S^1_0      ~.
\end{flalign}
These  equations are the   same as  $(34'), (37')$ and $(38')$ of  \cite{Kephart:1994uy}, when we set
\begin{flalign}  \nonumber  
&K^{0}_0={m_a^2\beta^3 \over 6\sqrt\pi }=N^{1}_0 ~,~  K^{01}_{000}={m_a^2\beta^3 \over 12\pi } ~, \\ \nonumber
&B^{1}_0={m_a^2\beta^2 \over 4\sqrt\pi } (1+ {2\beta \over 3}) ~,~ N^{1}_0={m_a^2\beta^3 \over 6\sqrt\pi }  ~,~
S^{1}_0={m_a^2\beta^2 \over 4\sqrt\pi }  ~,  \\ \nonumber
&n^a_{00}(t) = { n_a(t) \over 2\sqrt\pi }  ~,~ n^\gamma_{00}(t) = { n_\gamma(t) \over 2\sqrt\pi } ~.
\end{flalign}

The second case is
\begin{flalign}  \nonumber   \frac{ 3 }{  \beta^2} \sum_{l'}  n^a_{l'0}(t)   K^{0}_{l'} - 2\sum_{l'} n^\gamma_{l'0}(t)   N^1_{l'}=0 ~. \end{flalign}
This solution to eq. (\ref{A},\ref{B},\ref{C}) means that the effect of half of stimulated decay exactly cancels the effect of photon annihilation back into normal axion. This also expresses the $l=0$, $m=0$ component of axion number density as a sum of components of photon number density,
\begin{flalign}  \nonumber  
&     n^a_{00}(t)    = \frac{ 2 \beta^2}{ 3 K^{0}_{0} }\sum_{l'} n^\gamma_{l'0}(t)   N^1_{l'} ~.
\end{flalign}
Evolution equation \eqref{normal axion number density components equations} for $l=0$, $m=0$ component then becomes
\begin{flalign}  \nonumber  
 \frac{d n^a_{00}(t) }{dt}        =   - \frac{ 6  \Gamma_{a} \sqrt\pi  }{  m_a^2 \beta^3  }    [      K^{0}_0   n^a_{00}(t)  
                                                        + \frac{ 16\pi^3 }{ m_a^3 \beta}  \sum_{l''}   n^a_{00}(t)  n^\gamma_{l''0}(t)   K^{01}_{0l''0}   ] ~. 
\end{flalign}
Evolution equations \eqref{photon number density components equations} for any $l\neq0$, $m\neq0$ components of photon number density in this second case is
\begin{flalign}  \nonumber  
 \frac{d n^\gamma_{lm}(t) }{dt}        =     - \frac{ 64\pi^3 \Gamma_{a} }{ m_a^5 \beta^2  } n^\gamma_{lm}(t)  \sum_{l'} n^\gamma_{l'0}(t)   S^1_{l'}             - \frac{3c}{2R}n^\gamma_{lm}(t) ~.
\end{flalign}
Photon annihilation back into sterile axion and surface loss are the only mechanisms that contribute to $n^\gamma_{lm}(t)$($l\neq0$, $m\neq0$) modes. There is no source of decay providing photons to $l\neq0$, $m\neq0$ number density components. Therefore $n^\gamma_{lm}(t)$($l\neq0$, $m\neq0$) modes are expected to vanish quickly. The majority of photons would be in $n^\gamma_{00}(t)$ component.\\

In both cases, either $n^\gamma_{lm}(t)=0$ or $\frac{ 3 }{  \beta^2} \sum_{l'}  n^a_{l'0}(t)   K^{0}_{l'} - 2\sum_{l'} n^\gamma_{l'0}(t)   N^1_{l'}=0$, if the axions are locked in $Y_{00}$ momentum state, then so would be the photons (or at least predominantly in the second case).

\subsection{$Y_{20}$ momentum distribution}
In this configuration  the direction of momentum of axions is preferentially parallel to the polar axis.   It  describes the movement of axions between northern and southern hemispheres
\begin{flalign}  \nonumber  
n_a(t, \vec{r}, \Omega_p) &=\sum_{lm}n^a_{lm}(t)  \, \Theta(R-r) Y_{lm}(\Omega_p)       \\ \nonumber
                                                      &=n^a_{20}(t) \, \Theta(R-r) Y_{20}(\Omega_p)                  ~.
\end{flalign}
$n^a_{20}(t)$ is the only nonzero axion number density component, 
\begin{flalign}  \nonumber  n^a_{lm}(t)=0  \quad (l\neq 2 \quad m\neq0) ~. \end{flalign}
The evolution equation \eqref{normal axion number density components equations} for $n^a_{20}(t)$ becomes
\begin{flalign}  \nonumber  
 \frac{d n^a_{20}(t) }{dt}        =&   - \frac{  16\pi^3 \Gamma_{a} }{  m_a^5 \beta^2 } n^\gamma_{20}(t)                                                                                                                                                                                                                                                                                
                                                             [ \frac{ 3 }{  \beta^2}  n^a_{20}(t)   K^{0}_2 - 2\sum_{l'} n^\gamma_{l'0}(t)   N^1_{l'}  ]   \neq 0    ~.
\end{flalign}
At the same time, except for $n^a_{00}(t)$ and $n^a_{20}(t)$, evolution equation \eqref{normal axion number density components equations} becomes
\begin{flalign}  \nonumber  
 \frac{d n^a_{lm}(t) }{dt}        &=   - \frac{  16\pi^3 \Gamma_{a} }{  m_a^5 \beta^2  }    n^\gamma_{lm}(t)                                                                                                                                                                                                                                                                                
                                                             [ \frac{ 3 }{  \beta^2}   n^a_{20}(t)   K^{0}_2 - 2\sum_{l'} n^\gamma_{l'0}(t)   N^1_{l'}  ]    \\ \nonumber
&= 0   \quad (lm\neq 00,20)   ~.
\end{flalign}
Comparing these two equations, we have
\begin{flalign}  \nonumber  n^\gamma_{lm}(t)=0 \quad (lm\neq 00,20)  ~.\end{flalign}
This shows that there are only two nonzero components for photon number density, $n^\gamma_{00}(t)$ and $n^\gamma_{20}(t)$.\\

Since the $n^a_{00}(t)$ component of axion number density is 0, evolution equation \eqref{normal axion number density components equations} reduces to
\begin{flalign}  \nonumber  
\frac{d n^a_{00}(t) }{dt} =& 0     =   - \frac{   \Gamma_{a}  }{  m_a^2 \beta^2  }   \{  \sqrt\pi  [  \frac{6 }{ \beta}   K^{0}_2   n^a_{20}(t)   \\ \nonumber 
                                                      &  + \frac{ 96\pi^3 }{ m_a^3 \beta^2}  \sum_{l''}   n^a_{20}(t)  n^\gamma_{l''0}(t)   K^{01}_{2l''0}   ]   \\ \nonumber
                                                      &  + \frac{ 16\pi^3 n^\gamma_{00}(t)}{ m_a^3 }      [ \frac{ 3 }{  \beta^2}   n^a_{20}(t)   K^{0}_{2} - 2\sum_{l'} n^\gamma_{l'0}(t)   N^1_{l'}  ] \}      ~.
\end{flalign}
This can be used to simplify the photon evolution equation \eqref{photon number density components equations} for component $n^\gamma_{00}(t)$,
\begin{flalign}  \nonumber  
 \frac{d n^\gamma_{00}(t) }{dt}        = &   \frac{  2 \Gamma_{a} }{  m_a^2 \beta^2  }   \{  \sqrt\pi  [  \frac{6 }{ \beta}   K^{0}_2   n^a_{20}(t)   
                                                        + \frac{ 96\pi^3 }{ m_a^3 \beta^2}  \sum_{l''}   n^a_{20}(t)  n^\gamma_{l''0}(t)   K^{01}_{2l''0}  ]   \\ \nonumber
                                                      &  + \frac{ 16\pi^3 n^\gamma_{00}(t)}{ m_a^3 }      [ \frac{ 3 }{  \beta^2}   n^a_{20}(t)   K^{0}_2 - 2\sum_{l'} n^\gamma_{l'0}(t)   B^1_{l'}  ] \}     
                                                      - \frac{3c}{2R}n^\gamma_{00}(t)  \\ \nonumber
        =&  -  \frac{  64\pi^3 \Gamma_{a} }{  m_a^5 \beta^2  }      n^\gamma_{00}(t)   
  \sum_{l'} n^\gamma_{l'0}(t)   S^1_{l'}        - \frac{3c}{2R}n^\gamma_{00}(t) ~.
\end{flalign}
The $n^\gamma_{00}(t)$ component is expected to vanish quickly since the only contributions to this mode are due to back reaction to sterile axions and surface losses. Meanwhile, the photon component $n^\gamma_{20}(t)$ evolves according to equation \eqref{photon number density components equations},
\begin{flalign}  \nonumber  
\frac{d n^\gamma_{20}(t) }{dt}   =&    \frac{  32 \pi^3 \Gamma_{a} }{  m_a^2 \beta^5  }    n^\gamma_{20}(t)     [ \frac{ 3 }{  \beta^2}   n^a_{20}(t)   K^{0}_{2} - 2\sum_{l'} n^\gamma_{l'0}(t)   B^1_{l'}  ]       \\ \nonumber
                                                      &    - \frac{3c}{2R}n^\gamma_{20}(t)  ~.
\end{flalign}
If axions were locked in a $Y_{20}$ momentum state, the majority of photons would be expected to be in this state, which means more photons would travel in the direction that is to some extent parallel to the polar axis. 
However, this is not an acceptable physical number density for particles because $Y_{ 20 }$ is negative in some regions, but in the appropriate combinations with $Y_{ 00 }$,  the total number density can be positive.

\subsection{$  Y_1^{\pm1*} Y_1^{\pm1}\sim\sin^2\theta$ momentum distribution}
This configuration has the direction of momentum of axions preferentially parallel to the equatorial plane, and is a good approximation to the case of superradiant axions near a Kerr black hole \cite{Rosa:2017ury}. It  describes axions  rotating around a polar axis. The axion number density has two nonzero components, $n^a_{00}$ and $n^a_{20}$.
\begin{flalign}  \nonumber  
&n_a(t, \vec{r}, \Omega_p)   \\ \nonumber =& \sum_{lm}n^a_{lm}(t)   Y_{lm}(\Omega_p) \, \Theta(R-r)  \\ \nonumber
                                             =& n^a(t)  \sin^2\theta_p    \, \Theta(R-r)                                  \\ \nonumber
                                            =& n^a(t)  \frac{4\sqrt\pi}{3} [ Y_{00}(\Omega_p)-\frac{1}{\sqrt5}Y_{20}(\Omega_p) ]    \, \Theta(R-r) ~.
\end{flalign}
To maintain the $\sin^2\theta$ distribution, there is a relation between these two nonzero components,
\begin{flalign}  \label{Example-sin^2-axion00-20 relation}   n^a_{20}(t)= -\frac{1}{\sqrt5} n^a_{00}(t)   ~.  \end{flalign}

Evolution equation \eqref{normal axion number density components equations} for $n^a_{20}(t)$ becomes
\begin{flalign}  \label{Example-sin^2-axion20}
 \frac{d n^a_{20}(t) }{dt}        =&   - \frac{  16\pi^3 \Gamma_{a} }{  m_a^5 \beta^2 } n^\gamma_{20}(t)                                                                                                                                                                                                                                                                                
                                                             [ \frac{ 3 }{  \beta^2}  \sum_{l'}  n^a_{l'0}(t)   K^{0}_{l'} - 2\sum_{l'} n^\gamma_{l'0}(t)   N^1_{l'}  ]   \neq 0    ~,
\\  \nonumber 
\end{flalign}
while for components other than $n^a_{00}(t)$ and $n^a_{20}(t)$, evolution equation \eqref{normal axion number density components equations} becomes
\begin{flalign}  \nonumber  
 \frac{d n^a_{lm}(t) }{dt}        &=   - \frac{  16\pi^3 \Gamma_{a} }{  m_a^5 \beta^2  }    n^\gamma_{lm}(t)                                                                                                                                                                                                                                                                                
                                                             [ \frac{ 3 }{  \beta^2}   \sum_{l'}  n^a_{l'0}(t)   K^{0}_{l'} - 2\sum_{l'} n^\gamma_{l'0}(t)   N^1_{l'}  ]    \\ \nonumber
&= 0   \quad (lm\neq 00,20)   ~.
\end{flalign}
Comparing these two equations, we have
\begin{flalign}  \nonumber  n^\gamma_{lm}(t)=0 \quad (lm\neq 00,20)  ~.\end{flalign}
This shows that there are only two nonzero components for photon number density, $n^\gamma_{00}(t)$ and $n^\gamma_{20}(t)$.\\

In Appendix D, we find that the quantity $n^\gamma_{00}(t) + \sqrt5 n^\gamma_{20}(t)$ evolves based on the following equation,
\begin{flalign}  \nonumber  
&  \frac{d  }{dt}           [ n^\gamma_{00}(t) + \sqrt5 n^\gamma_{20}(t)   ]      \\ \nonumber
                                                           =&   -  [  \frac{ 64\pi^3 \Gamma_{a} }{ m_a^5 \beta^2 }   \sum_{l'} n^\gamma_{l'0}(t)   S^1_{l'}  + \frac{3c}{2R}   ]  [ n^\gamma_{00}(t) + \sqrt5 n^\gamma_{20}(t)   ]  ~.
\end{flalign}
Therefore $n^\gamma_{00}(t) + \sqrt5 n^\gamma_{20}(t) = 0$ is a possible solution to this differential equation although it is not the uniquely solution, which indicates that axions of $\sin^2\theta$ momentem distribution can still generate photons of $\sin^2\theta$ momentem distribution. However, since only photon annihilation back into sterile axion and surface loss contribute to the change of the quantity $n^\gamma_{00}(t) + \sqrt5 n^\gamma_{20}(t)$, we expect,
\begin{flalign}  \nonumber  n^\gamma_{00}(t) + \sqrt5 n^\gamma_{20}(t) \approx 0  ~. \end{flalign}
So approximately, decay from $\sin^2\theta$ momentum distribution of axions results in a similar directional profile of photons.
\section{Summary and conclusions}
Equations (13--15) contain the main result of this work, where
we have modeled clusters of axions with spherically symmetric spacial but arbitrary momentum distributions and have studied the directional profile of photos produced in their evolution through spontaneous and stimulated axion decay    via the process $a \rightarrow \gamma  \gamma$. These results can be used in  situations where astrophysical axions cluster. Axions are a prime candidate for dark matter and clustering can be due to any number of reasons ranging from primodial density perturbations to superradience around black holes. Three specific examples were presented, one with spherical symmetry. to make contact with previous work, and two other typical but simple examples without spherical symmetry. It is straightforward to use equations (13--15) to model any cluster of axions with spherically symmetric spacial but arbitrary momentum distributions so we believe our results will have wide application. In future work we intend to relax the requirement of spherical symmetry on the spacial distribution.

\section{Acknowledgments} This work  was supported by US DOE grant DE-SC0019235.

\section{Appendix A: kinematics}
Here we collect some useful kinematic relations needed in the text.
\subsection{$  \delta [|\vec{p}-\vec{k}|-(p^0-k) ] $ conversion   }
Choose the $z$-axis to align with photon momentum $\vec{k}$, then  
\begin{flalign} \nonumber
&  \cos\theta_p=\frac{\vec{p} \cdot \vec{k}}{ p k }  \\  \nonumber
&  (\vec{p}-\vec{k})^2  = p^2 + k^2 - 2pk\cos\theta_p \\  \nonumber
=&  (p^0)^2-m_a^2+k^2-2k\sqrt{(p^0)^2-m_a^2}\cos\theta_p    ~.
\end{flalign}
Now let $g(\cos\theta_p)=|  \vec{p}-\vec{k} |$, to find
\begin{flalign} \nonumber
&  \delta [|\vec{p}-\vec{k}|-(p^0-k) ]  = \delta [ g(\cos\theta_p)  -  (p^0-k) ]   \\  \nonumber
=&  \delta\bigg[\sqrt{(p^0)^2-m_a^2+k^2-2k\sqrt{(p^0)^2-m_a^2}\cos\theta_p}-(p^0-k)\bigg]  \\  \nonumber
=&  \delta ( \cos\theta_p - \cos\theta_{p0} )  \times \bigg|\frac{d\cos\theta_p}{dg(\cos\theta_p)}\bigg|_{\cos\theta_p=\cos\theta_{p0}}  ~,
\end{flalign}
where $\cos\theta_{p0}$ is the root of the equation $ g(\cos\theta_p)  -  (p^0-k)  = 0 $. Clearly
\begin{flalign} \nonumber
& g(\cos\theta_{p0})  -  (p^0-k)  = 0 \quad\Leftrightarrow\quad (\vec{p}-\vec{k})^2=(p^0-k)^2  \\  \nonumber
& p^2 + k^2 -  2 pk\cos\theta_{p0}  = (p^0)^2 + k^2 - 2p^0 k   \\  \label{cos-theta-p0}
&   \Rightarrow \cos\theta_{p0} = \frac{2p^0 k - m_a^2}{2k\sqrt{(p^0)^2-m_a^2}}      ~.
\end{flalign}
The derivative at $\cos\theta_{p0}$ can also be calculated, 
\begin{flalign} \nonumber
\frac{dg(\cos\theta_p)}{d\cos\theta_p} &=  \frac{-k\sqrt{(p^0)^2-m_a^2}}{g(\cos\theta_p)+p^0-k}   \\  \nonumber
\end{flalign}
which can be evaluated to give
\begin{flalign} \nonumber
  \bigg|\frac{d\cos\theta_p}{dg(\cos\theta_p)}\bigg|_{\cos\theta_p=\cos\theta_{p0}}  &= \frac{p^0-k}{k\sqrt{(p^0)^2-m_a^2}}  \\  \nonumber
~.
\end{flalign}
Substituting the root and the derivative, and $\delta [ |\vec{p}-\vec{k}|-(p^0-k) ]$  is converted to a $\delta$ function with respect to $\theta_p$,
\begin{flalign} \nonumber
 &  \delta |\vec{p}-\vec{k}|-(p^0-k) ]  
=   \frac{p^0-k}{k\sqrt{(p^0)^2-m_a^2}}    \delta\bigg[ \cos\theta_p - \frac{2p^0 k - m_a^2}{2k\sqrt{(p^0)^2-m_a^2}} \bigg]  \\  
 \nonumber
 \end{flalign}
 or
 \begin{flalign} \nonumber
=   \frac{p^0-k}{k\sqrt{(p^0)^2-m_a^2}  \sin\theta_{p0}}   \delta( \theta_p - \theta_{p0}  )    ~.
\end{flalign}

 Then the common way to obtain various  trigonometric quantities about $\vec{p}$ is through relations for the angles given in the appendix
\begin{flalign}  \nonumber  
\cos\theta_p &=   \frac{\vec{p} \cdot \vec{k}}{pk}     ~,      \\ \nonumber
\sin\theta_p\cos\phi_p  &=   \frac{\vec{p} \cdot \vec{e}_x}{p}   ~,~    \sin\theta_p\sin\phi_p  =   \frac{\vec{p} \cdot \vec{e}_y}{p}      ~.  
\end{flalign}
Similarly,  trigonometric quantities about $\vec{p}-\vec{k}$ can be obtained via following relations,
\begin{flalign}  \nonumber  
&  \cos\theta_{\vec{p}-\vec{k}}     
=   \frac{ (\vec{p}-\vec{k}) \cdot \vec{k}}{ |\vec{p}-\vec{k}| k}          
=   \frac{ p\cos\theta_p - k }{  \sqrt{ p^2 + k^2 -2pk\cos\theta_p }  }      ~,    \\ \nonumber  
&  \sin\theta_{\vec{p}-\vec{k}}   
=   \sqrt{1-\cos^2\theta_{\vec{p}-\vec{k}}} =   \frac{   p\sin\theta_p   }{  \sqrt{ p^2 + k^2 -2pk\cos\theta_p }  }        ~,    
\end{flalign}
\begin{flalign}  \nonumber  
\sin\theta_{\vec{p}-\vec{k}}\cos\phi_{\vec{p}-\vec{k}}  &=   \frac{ (\vec{p}-\vec{k}) \cdot \vec{e}_x}{ |\vec{p}-\vec{k}| } 
                                                                                              =   \frac{ \vec{p} \cdot \vec{e}_x}{ |\vec{p}-\vec{k}| }  \\ \nonumber
                                                                                            &= \frac{ p\sin\theta_p\cos\phi_p }{ |\vec{p}-\vec{k}| }    
                                                                                              =  \sin\theta_{\vec{p}-\vec{k}} \cos\phi_p    ~,  \\ \nonumber
\sin\theta_{\vec{p}-\vec{k}}\sin\phi_{\vec{p}-\vec{k}}  &=   \frac{ (\vec{p}-\vec{k}) \cdot \vec{e}_y}{ |\vec{p}-\vec{k}| } 
                                                                                              =   \frac{ \vec{p} \cdot \vec{e}_y}{ |\vec{p}-\vec{k}| }  \\ \nonumber
                                                                                            &= \frac{ p\sin\theta_p\sin\phi_p }{ |\vec{p}-\vec{k}| }    
                                                                                              =  \sin\theta_{\vec{p}-\vec{k}} \sin\phi_p     ~.
\end{flalign}
Thus $\phi_{\vec{p}-\vec{k}}$  and $\phi_p$ are equal,
\begin{flalign}  \nonumber  
\cos\phi_{\vec{p}-\vec{k}} =\cos\phi_p  \quad    \sin\phi_{\vec{p}-\vec{k}}=\sin\phi_p    
    \quad\Rightarrow \phi_{\vec{p}-\vec{k}}   =\phi_p ~.
\end{flalign}
Then the third $\phi_p$ integration from equation \eqref{10Kephart-2} can be carried out,
\begin{flalign}  \nonumber  
&\quad  \int   \sum_{lm}b_{lm}(p^0-k,t)Y_{lm}(\Omega_{\vec{p}-\vec{k}})   \,   d\phi_p      \\ \nonumber
                                         &=  \sum_{lm}b_{lm}(p^0-k,t)  \int   Y_{lm}(\theta_{\vec{p}-\vec{k}}, \phi_p)   \,   d\phi_p  \\ \nonumber
                                         &=  \sum_{lm}b_{lm}(p^0-k,t)  Y_{lm}(\theta_{\vec{p}-\vec{k}}, \phi_p) \times 2\pi \delta_{m0}  \\ \nonumber
                                         &=2\pi \sum_{l}b_{l0}(p^0-k,t)  Y_{l0}(\theta_{\vec{p}-\vec{k}})     ~.
\end{flalign}

\section{Appendix B: Relations between real and complex spherical harmonics}
 If $m' , m<0$, then we can write
\begin{flalign}  \nonumber  
&\  \int   Y_{l'm'}(\Omega_p)  Y_{lm}(\Omega_{\vec{p}-\vec{k}}) \,  d\phi_p    \\ \nonumber
                                 =& \int  \frac{i}{\sqrt 2}   [ Y_{l'}^{m'}(\Omega_p) -(-1)^{m'} Y_{l'}^{-m'}(\Omega_p) ]   \\ \nonumber
                                          &\quad   \times\frac{i}{\sqrt 2}  [ Y_{l}^{m}(\Omega_{\vec{p}-\vec{k}}) 
                                               -(-1)^{m} Y_{l}^{-m}(\Omega_{\vec{p}-\vec{k}}) ]    \,  d\phi_p    \\ \nonumber
                          \end{flalign}    
                          
              Expanding and using the property $ {Y_{l}^{m}}^*=(-1)^{m} Y_{l}^{-m}$  we find                          
                          \begin{flalign}  \nonumber 
                          &\  \int   Y_{l'm'}(\Omega_p)  Y_{lm}(\Omega_{\vec{p}-\vec{k}}) \,  d\phi_p    \\ \nonumber                     
                                 =& \int  \frac{1}{ 2}   [ Y_{l'}^{m'}(\Omega_p)  Y_{l}^{m}(\Omega_{\vec{p}-\vec{k}})^*
                                                                      +  Y_{l}^{m}(\Omega_{\vec{p}-\vec{k}})     Y_{l'}^{m'}(\Omega_p)^*   \\ \nonumber
                                          &\quad   -   Y_{l'}^{m'}(\Omega_p)   Y_{l}^{m}(\Omega_{\vec{p}-\vec{k}}) 
                                               - Y_{l}^{m}(\Omega_{\vec{p}-\vec{k}})^*  Y_{l'}^{m'}(\Omega_p)^*  ]    \,  d\phi_p     ~. \\ \nonumber
\end{flalign}
We are integrating over $\phi_p$, but since $\phi_{\vec{p}-\vec{k}} = \phi_p$, it doesn't matter that $\theta_{\vec{p}-\vec{k}} \neq \theta_p$ because $Y_{l}^{m}$ is of the form $P_l^m(\cos\theta) e^{im\phi}$. $Y_{l}^{m}(\theta, 0)=Y_{l}^{m}(\theta, \phi=0)$, which is a real function, and is used to denote the result from integration of complex spherical harmonics over azimuthal angle,
\begin{flalign}  \nonumber  
&\quad  \int   Y_{l'm'}(\Omega_p)  Y_{lm}(\Omega_{\vec{p}-\vec{k}}) \,  d\phi_p    \\ \nonumber
                                 &=   \frac{1}{ 2}   [ 2\pi\delta_{m'm}Y_{l'}^{m'}(\theta_p,0)  Y_{l}^{m}(\theta_{\vec{p}-\vec{k}},0)    \\ \nonumber
                                                             &+  2\pi\delta_{m'm}Y_{l}^{m}(\theta_{\vec{p}-\vec{k}},0)     Y_{l'}^{m'}(\theta_p,0)   ]   \\ \nonumber
                                 &=    2\pi\delta_{m'm}Y_{l'}^{m}(\theta_p,0)  Y_{l}^{m}(\theta_{\vec{p}-\vec{k}},0)    ~~(m' , m<0)  ~. 
\end{flalign}
This result actually is applicable to all cases of  signs of $m'$ and $m$, so we can drop the restriction and write simply
\begin{flalign}  \nonumber  
\int   Y_{l'm'}(\Omega_p)  Y_{lm}(\Omega_{\vec{p}-\vec{k}})    d\phi_p    =    2\pi\delta_{m'm}Y_{l'}^{m}(\theta_p,0)  Y_{l}^{m}(\theta_{\vec{p}-\vec{k}},0)    .
\end{flalign}

\vfill

\section{Appendix C: Momentum space angular integration}
 \subsection{Integration over $\phi_p$: $Y_{lm}(\Omega_p)$ and $Y_{lm}(\Omega_{\vec{p}-\vec{k}})$ }
The  $\phi_p$ integration of the first term in braces in eq. \eqref{10Kephart-2} can be calculated directly using eq. \eqref{f_a}. Due to the factor $e^{i m \phi}$ contained in $Y_{lm}$,  the integration is 0 unless $m=0$.
\begin{flalign}  \nonumber  
&  \int  f_a  ( \vec{p} )  \, d\phi_p    
=         2\pi  \sum_{l}a_{l0}(p,t)   Y_{l0}( \theta_p )       ~.
\end{flalign}
As stated before, $\vec{k}$ is chosen as the $z$-axis.%
\subsection{Integration over $\phi_p$: $Y_{l'm'}(\Omega_p)  Y_{lm}(\Omega_{\vec{p}-\vec{k}})$}
Again substituting the form of $f_a$ from eq. \eqref{f_a}, we see the  $\phi_p$ integration in the second term of eq. \eqref{10Kephart-2} has to be performed on $Y_{l'm'}(\Omega_p)  Y_{lm}(\Omega_{\vec{p}-\vec{k}})$, 
The calculation is easier if we convert from real back to the complex spherical harmonics. But the relation between real and complex spherical harmonics   depends on the signs of $m'$ and $m$.
Using the relations among spherical harmonics found in  Appendix B facilitates the   integration and we find,
\begin{flalign}  \nonumber  
&  \int  f_a  ( \vec{p} )  \sum_{lm}b_{lm}(p^0-k ,t)Y_{lm}(\Omega_{\vec{p}-\vec{k}}) \,  d\phi_p    \\ \nonumber
                                         =&  2\pi\sum_{l'lm}a_{l'm}(p,t) b_{lm}(p^0-k ,t) 
                                                            Y_{l'}^{m}(\theta_p,0)  Y_{l}^{m}(\theta_{\vec{p}-\vec{k}},0)    ~.    
\end{flalign}

The third integral is similar to the first.

Collecting terms 
after all $\phi_p$ integrations   eq. \eqref{10Kephart-2} becomes
\begin{flalign}  \nonumber
&   2k \frac{d f_\lambda ( \vec{k} )  }{dt}  =   \frac{ m_{a} \Gamma_{a} }{ \pi } 
                      \int  \frac{   d p^0 \,  \delta( \theta_p - \theta_{p0}  )    }{k  \sin\theta_{p0}}   \sin\theta_p d\theta_p   \\ \nonumber
                 &     \times \{    [ 1+ f_\lambda (\vec{k} ) ]   \times  2\pi  \sum_{l}a_{l0}(p,t)   Y_{l0}( \theta_p )  +  \\ \nonumber
                &    2\pi\sum_{l'lm}a_{l'm}(p,t) b_{lm}(p^0-k ,t) Y_{l'}^{m}(\theta_p,0)  Y_{l}^{m}(\theta_{\vec{p}-\vec{k}},0)   \\ \nonumber
                &     -   f_\lambda ( \vec{k} )  \times 2\pi \sum_{l}b_{l0}(p^0-k,t)  Y_{l0}(\theta_{\vec{p}-\vec{k}})  \}  ~.
\end{flalign}
Integration over $\theta_p$ is straightforward and leads to
\begin{flalign} \nonumber
&  k \frac{d f_\lambda ( \vec{k} )  }{dt}   =  m_{a} \Gamma_{a}
                      \int  \frac{   d p^0     }{k  }       \{    [ 1+ f_\lambda (\vec{k} ) ]      \sum_{l}a_{l0}(p,t)   Y_{l0}( \theta_{p0} ) + \\ \nonumber 
                &     \sum_{l'lm}a_{l'm}(p,t) b_{lm}(p^0-k ,t) Y_{l'}^{m}(\theta_{p0},0)     Y_{l}^{m}(\theta_{\vec{p}-\vec{k}}|_{\theta_p=\theta_{p0}},0)   \\ \nonumber
                &    -   f_\lambda ( \vec{k} )  \times   \sum_{l}b_{l0}(p^0-k,t)  Y_{l0}(\theta_{\vec{p}-\vec{k}}|_{\theta_p=\theta_{p0}})  \}  ~.
\end{flalign}

but $ Y_{l}^{m}(\theta_{\vec{p}-\vec{k}}, 0)$ and $Y_{l0}(\theta_{\vec{p}-\vec{k}})$ need to be evaluated at $\theta_p=\theta_{p0}$ since $\theta_{\vec{p}-\vec{k}}$ is an implicit function of $\theta_p$. $\cos\theta_p$ has been evaluated at $\theta_{p0}$ by \eqref{cos-theta-p0}, 
\begin{flalign}  \nonumber  
&  \cos\theta_p|_{\theta_p = \theta_{p0}} = \cos\theta_{p0}  = \frac{2p^0 k - m_a^2}{2k\sqrt{(p^0)^2-m_a^2}} ~.
 \end{flalign}
Note that $|\vec{p}-\vec{k}| =(p^0-k)$ is equivalent to $\theta_p = \theta_{p0}$ because that is how \eqref{cos-theta-p0} was derived. This gives us a relation between $\cos\theta_{\vec{p}-\vec{k}}$ and $\cos\theta_p$,
\begin{flalign}  \nonumber  
  \cos\theta_{\vec{p}-\vec{k}} =& \frac{ (\vec{p}-\vec{k}) \cdot \vec{k}}{ |\vec{p}-\vec{k}| k}   
                                                     = \frac{ p\cos\theta_p  - k }{ p^0 - k }   ~, \\ \nonumber
   \cos\theta_{\vec{p}-\vec{k}}|_{\theta_p = \theta_{p0}}  =&   \frac{ p^0 - \frac{m_a^2}{2k} - k }{ p^0-k } = 1 - \frac{m_a^2}{2k(p^0-k)}                 \\ \nonumber
     =&     \cos\theta_{\vec{p}-\vec{k}}|_{\theta_{\vec{p}-\vec{k}} = \theta_{p1}}   ~.    \end{flalign}
The last line suggests that $\cos\theta_{\vec{p}-\vec{k}}$ can also be directly evaluated at $\theta_{\vec{p}-\vec{k}}= \theta_{p1}$ if 
\begin{flalign}  \label{p1}
&\cos\theta_{p1} =  1 - \frac{m_a^2}{2k(p^0-k)}  ~.
\end{flalign}
Adopting this $\theta_{p1}$ notation, eq. \eqref{10Kephart-2} become
\begin{flalign}  \nonumber  
&  k \frac{d f_\lambda ( \vec{k} )  }{dt}    \\ \nonumber 
=&   m_{a} \Gamma_{a}    \int  \frac{   d p^0     }{k  }  ~     \{    [ 1+ f_\lambda (\vec{k} ) ]   
       \sum_{l}a_{l0}(p,t)   Y_{l0}( \theta_{p0} )  +  \\ \nonumber 
                &    \sum_{l'lm}a_{l'm}(p,t) b_{lm}(p^0-k ,t) Y_{l'}^{m}(\theta_{p0},0)  Y_{l}^{m}(\theta_{p1},0)   \\ \label{10Kephart-3}
                &     -   f_\lambda ( \vec{k} )  \times   \sum_{l}b_{l0}(p^0-k,t)  Y_{l0}(\theta_{p1})  \}  ~.
\end{flalign}

\subsection{Integration over $k_1$ and $k$  }
Equation \eqref{occupation number components equations} needs to be integrated over $k$ to obtain an equation for the components of number density. To do this we first rewrite eq. \eqref{occupation number components equations} as
\begin{flalign}  \label{integration over occupation number equation}  
& \int \frac{k^2dk}{(2\pi)^3}  \frac{ d b_{lm}(k,t) }{ dt }     \\ \nonumber   
=&  \int \frac{k^2dk}{(2\pi)^3}  \frac{  m_{a} \Gamma_{a} }{k^2  }   \int_{\frac{m_a^2}{4k}} dk_1   \{ \delta_{l0}\delta_{m0} 2\sqrt\pi                              
                   [ \sum_{l'}  a_{l'0}( p,t)   Y_{l'0}( \theta_{p0} )       \\ \nonumber  
                                                             &+ \sum_{l'l''m'}  a_{l'm'}(p,t) b_{l''m'}(k_1 ,t) Y_{l'}^{m'}(\theta_{p0},0)  Y_{l''}^{m'}(\theta_{p1},0) ]   \\ \nonumber
                                                      &  + b_{lm}(k,t)      [ \sum_{l'}a_{l'0}(p,t)   Y_{l'0}( \theta_{p0} ) -\sum_{l'}b_{l'0}(k_1,t)  Y_{l'0}(\theta_{p1})  ] \}   .
\end{flalign}
Apart from writing down step function $\Theta(R-r)$,  integrations are carried over  \eqref{occupation number components equations} term by term in the following calculation,
\begin{flalign}  \nonumber  
&  \int \frac{k^2dk}{(2\pi)^3} \frac{ d b_{lm}(k,t) }{ dt }  \\ \nonumber
=& \frac{ d b_{lm}(t) }{ dt }   \int \Theta(k_+-k) \Theta(k-k_-) \, \frac{k^2dk}{(2\pi)^3}   = \frac{d n^\lambda_{lm}(t) }{dt}    ~.
\end{flalign}
The integration on the RHS of \eqref{integration over occupation number equation} is over $k$ and $k_1$.  The following calculations omit writing down common factor $  m_{a} \Gamma_{a}  \over (2\pi)^3 $. The first term on the RHS of \eqref{integration over occupation number equation} is
\begin{flalign}  \nonumber  
&  \int dk \int_{\frac{m_a^2}{4k}} dk_1 \sum_{l'}  a_{l'0}( p,t)   Y_{l'0}( \theta_{p0} ) \\ \nonumber
=&\sum_{l'} a_{l'0}(t)   \int dk  \int_{\frac{m_a^2}{4k}} dk_1  \Theta(p_{\mbox{\tiny max}}-p)  Y_{l'0}( \theta_{p0} )  \\ \nonumber
=&\sum_{l'} a_{l'0}(t)   \int dk  \int_{\frac{m_a^2}{4k}}^{m_a\gamma-k} dk_1  \sqrt{\frac{2l'+1}{4\pi}}    P_{l'}^0( \cos\theta_{p0} )  \\ \nonumber
=&\sum_{l'} a_{l'0}(t)   K^{0}_{l'}  ~,
\end{flalign}
where $K^{0}_{l'}$ are defined to be constant coefficients describing spontaneous and half of stimulated decay, and are related to associated Legendre polynomials $P_{l'}^0$.
\begin{flalign}  \nonumber  
K^{0}_{l'}=\sqrt{\frac{2l'+1}{4\pi}}     \int dk  \int_{\frac{m_a^2}{4k}}^{m_a\gamma-k}   dk_1    P_{l'}^0[ \frac{2(k_1+k) k - m_a^2}{2k\sqrt{(k_1+k)^2-m_a^2}}]   .
\end{flalign}
The second term on the RHS of \eqref{occupation number components equations} can be treated similarly,
\begin{flalign}  \nonumber  
&\int dk  \int_{\frac{m_a^2}{4k}} dk_1 \sum_{l'l''m'}  a_{l'm'}(p,t) b_{l''m'}(k_1 ,t)    \\ \nonumber
          &\qquad  \times  Y_{l'}^{m'}(\theta_{p0},0)  Y_{l''}^{m'}(\theta_{p1},0)  \\ \nonumber
=&\sum_{l'l''m'} a_{l'm'}(t) b_{l''m'}(t) \int dk \int_{\frac{m_a^2}{4k}} dk_1   \Theta(p_{\mbox{\tiny max}}-p)    \\ \nonumber
          &\qquad  \times  Y_{l'}^{m'}(\theta_{p0},0)  Y_{l''}^{m'}(\theta_{p1},0)    \Theta(k_+-k_1) \Theta(k_1-k_-) \\ \nonumber
=&\sum_{l'l''m'} a_{l'm'}(t) b_{l''m'}(t) \int dk  \int_{\frac{m_a^2}{4k}}^{m_a\gamma-k} dk_1   \\ \nonumber
          &\qquad  \times  Y_{l'}^{m'}(\theta_{p0},0)  Y_{l''}^{m'}(\theta_{p1},0)  \\ \nonumber
=&  \sum_{l'l''m'} a_{l'm'}(t) b_{l''m'}(t)  K^{01}_{l'l''m'}  ~,
\end{flalign}
where $K^{01}_{l'l''m'}$ are constant coefficients describing the other half of stimulated decay and are related to associated Legendre polynomials $ P_{l'}^{m'}$, $P_{l''}^{m'}$.
\begin{flalign}  \nonumber  
K^{01}_{l'l''m'} =&  \int dk  \int_{\frac{m_a^2}{4k}}^{m_a\gamma-k} dk_1  Y_{l'}^{m'}(\theta_{p0},0)  Y_{l''}^{m'}(\theta_{p1},0) \\ \nonumber
=& \sqrt{\frac{(2l'+1)(l'-m')!}{4\pi(l'+m')!}}  \sqrt{\frac{(2l''+1)(l''-m')!}{4\pi(l''+m')!}}  \\ \nonumber
&    \times  \int dk  \int_{\frac{m_a^2}{4k}}^{m_a\gamma-k} dk_1  \times  \\ \nonumber
&      P_{l'}^{m'}[ \frac{2(k_1+k) k - m_a^2}{2k\sqrt{(k_1+k)^2-m_a^2}}] \   P_{l''}^{m'}[ 1-\frac{m_a^2}{2kk_1} ] .
\end{flalign}
The third term on the RHS of \eqref{integration over occupation number equation} is simplified by using newly defined coefficients $K^{0}_{l'}$.
\begin{flalign}  \nonumber  
&  \int dk \int_{\frac{m_a^2}{4k}} dk_1 b_{lm}(k,t) \sum_{l'}  a_{l'0}( p,t)   Y_{l'0}( \theta_{p0} ) \\ \nonumber
=& b_{lm}(t)\sum_{l'} a_{l'0}(t)   K^{0}_{l'} ~.
\end{flalign}
The last term on the RHS of \eqref{integration over occupation number equation} is
\begin{flalign}  \nonumber  
&\int dk  \int_{\frac{m_a^2}{4k}} dk_1 b_{lm}(k,t)  \sum_{l'}  b_{l'0}( k_1,t)   Y_{l'0}( \theta_{p1} ) \\ \nonumber
=&b_{lm}(t)\sum_{l'} b_{l'0}(t)  \int dk \int_{\frac{m_a^2}{4k}} dk_1  \\ \nonumber
   &\times \Theta(k_+-k_1) \Theta(k_1-k_-)  Y_{l'0}( \theta_{p1} )  \\ \nonumber
=&b_{lm}(t)\sum_{l'} b_{l'0}(t)  \int dk \int_{\frac{m_a^2}{4k}}^{k_+}  dk_1   Y_{l'0}( \theta_{p1} )  \\ \nonumber
=&b_{lm}(t)\sum_{l'} b_{l'0}(t)  B^1_{l'}  ~,
\end{flalign}
where $B^1_{l'}$ are constant coefficients describing  back reaction of photons and are related to associated Legendre polynomials $P_{l'}^0$.
\begin{flalign}  \nonumber  
 B^1_{l'} &=  \int dk   \int_{\frac{m_a^2}{4k}}^{k_+}  dk_1     Y_{l'0}( \theta_{p1} )  \\ \nonumber
&= \sqrt{\frac{2l'+1}{4\pi}}   \int dk   \int_{\frac{m_a^2}{4k}}^{k_+}  dk_1   P_{l'}^0[ 1-\frac{m_a^2}{2kk_1}  ]  ~.
\end{flalign}
However, back reaction can produce sterile axions with energy higher than $m_a\gamma$ which need to be excluded from axion number counting since we focus on nonrelativistic axions. Dividing the integral inteval $[\frac{m_a^2}{4k}, k_+]$ into two parts gives two sets of constant coefficients $N^1_{l'}$ and $S^1_{l'}$ describing back reactions produce normal axions and sterile axions, respectively, where $  N^1_{l'} + S^1_{l'} = B^1_{l'}  $ .
\begin{flalign}  \nonumber  
N^1_{l'} &=  \int dk   \int_{\frac{m_a^2}{4k}}^{m_a\gamma-k}  dk_1     Y_{l'0}( \theta_{p1} )  \\ \nonumber
&= \sqrt{\frac{2l'+1}{4\pi}}   \int dk   \int_{\frac{m_a^2}{4k}}^{m_a\gamma-k}  dk_1   P_{l'}^0[ 1-\frac{m_a^2}{2kk_1}  ]  ~.
\end{flalign}
\begin{flalign}  \nonumber  
S^1_{l'} &=  \int dk   \int_{m_a\gamma-k}^{k_+}  dk_1     Y_{l'0}( \theta_{p1} )  \\ \nonumber
&= \sqrt{\frac{2l'+1}{4\pi}}   \int dk   \int_{m_a\gamma-k}^{k_+}  dk_1   P_{l'}^0[ 1-\frac{m_a^2}{2kk_1}  ]  ~.
\end{flalign}

\subsection{Evolution equation for number density components}
Plugging the previously defined constant coefficients into \eqref{integration over occupation number equation}, we have differential equations about the photon number density components for helicity state $\lambda$.
\begin{flalign}  \nonumber  
& \frac{d n^\lambda_{lm}(t) }{dt}        =    \frac{  m_{a} \Gamma_{a} }{ (2\pi)^3  }   \{ \delta_{l0}\delta_{m0} 2\sqrt\pi  [  \sum_{l'} a_{l'0}(t)   K^{0}_{l'}   \\ \nonumber 
                                                      &  + \sum_{l'l''m'} a_{l'm'}(t) b_{l''m'}(t)  K^{01}_{l'l''m'}   ]   \\ \nonumber
                                                      &  + b_{lm}(t)  \times   [ \sum_{l'} a_{l'0}(t)   K^{0}_{l'} - \sum_{l'} b_{l'0}(t)  B^1_{l'}  ] \}    ~.
\end{flalign}
The components of occupation number $a_{lm}$ and $b_{lm}$ can be replaced by the components of number desintity, using \eqref{axion-an} and \eqref{photon-bn}.
\begin{flalign}  \nonumber  
& \frac{d n^\lambda_{lm}(t) }{dt}        =    \frac{  m_{a} \Gamma_{a} }{ (2\pi)^3  }   \{ \delta_{l0}\delta_{m0} 2\sqrt\pi  [  \sum_{l'} \frac{24\pi^3 n^a_{l'0}(t)}{ m_a^3 \beta^3}    K^{0}_{l'}   \\ \nonumber 
                                                      &  + \sum_{l'l''m'} \frac{24\pi^3 n^a_{l'm'}(t)}{ m_a^3 \beta^3}    \frac{32\pi^3 n^\lambda_{l''m'}(t)}{ m_a^3 \beta}   K^{01}_{l'l''m'}   ]   \\ \nonumber
                                                      &  + \frac{32\pi^3 n^\lambda_{lm}(t)}{ m_a^3 \beta}   \times   [ \sum_{l'} \frac{24\pi^3 n^a_{l'0}(t)}{ m_a^3 \beta^3}    K^{0}_{l'} - \sum_{l'} \frac{32\pi^3 n^\lambda_{l'0}(t)}{ m_a^3 \beta}   B^1_{l'}  ] \}    ~.
\end{flalign}
Combining common factors simplifies this to
\begin{flalign}  \nonumber  
& \frac{d n^\lambda_{lm}(t) }{dt}        =    \frac{  2 \Gamma_{a} }{  m_a^2 \beta^2  }   \{ \delta_{l0}\delta_{m0} \sqrt\pi  [  \sum_{l'} \frac{3 }{ \beta}   K^{0}_{l'}   n^a_{l'0}(t)   \\ \nonumber 
                                                      &  + \frac{ 96\pi^3 }{ m_a^3 \beta^2}  \sum_{l'l''m'}   n^a_{l'm'}(t)  n^\lambda_{l''m'}(t)   K^{01}_{l'l''m'}   ]   \\ \nonumber
                                                      &  + \frac{ 16\pi^3 n^\lambda_{lm}(t)}{ m_a^3 }   \times   [ \frac{ 3 }{  \beta^2} \sum_{l'}  n^a_{l'0}(t)   K^{0}_{l'} - 4\sum_{l'} n^\lambda_{l'0}(t)   B^1_{l'}  ] \}     ~.
\end{flalign}
We assume that the same component of number density of photon of each helicity state are the same, meaning that
\begin{flalign}  \nonumber  
n^+_{lm}(t) = n^-_{lm}(t)  ~,~  n^\gamma_{lm}(t)=2n^\lambda_{lm}(t)  ~.
\end{flalign}
This will give us the evolution equations for the individual components of number density of photon\eqref{photon number density components equations}, normal axion\eqref{normal axion number density components equations}, and sterile axion\eqref{sterile axion number density components equations}.
\begin{flalign}  \tag{\ref{photon number density components equations}  }
& \frac{d n^\gamma_{lm}(t) }{dt}        =    \frac{  2 \Gamma_{a} }{  m_a^2 \beta^2  }   \{ \delta_{l0}\delta_{m0} \sqrt\pi  [  \sum_{l'} \frac{6 }{ \beta}   K^{0}_{l'}   n^a_{l'0}(t)   \\ \nonumber 
                                                      &  + \frac{ 96\pi^3 }{ m_a^3 \beta^2}  \sum_{l'l''m'}   n^a_{l'm'}(t)  n^\gamma_{l''m'}(t)   K^{01}_{l'l''m'}   ]   \\ \nonumber
                                                      &  + \frac{ 16\pi^3 n^\gamma_{lm}(t)}{ m_a^3 }   \times   [ \frac{ 3 }{  \beta^2} \sum_{l'}  n^a_{l'0}(t)   K^{0}_{l'} - 2\sum_{l'} n^\gamma_{l'0}(t)   B^1_{l'}  ] \}      \\ \nonumber
                                                      &- \frac{3c}{2R}n^\gamma_{lm}(t)  ~.
\end{flalign}
\begin{flalign}  \tag{\ref{normal axion number density components equations}  }
& \frac{d n^a_{lm}(t) }{dt}        =   - \frac{   \Gamma_{a} }{  m_a^2 \beta^2  }   \{ \delta_{l0}\delta_{m0} \sqrt\pi  [  \sum_{l'} \frac{6 }{ \beta}   K^{0}_{l'}   n^a_{l'0}(t)   \\ \nonumber 
                                                      &  + \frac{ 96\pi^3 }{ m_a^3 \beta^2}  \sum_{l'l''m'}   n^a_{l'm'}(t)  n^\gamma_{l''m'}(t)   K^{01}_{l'l''m'}   ]   \\ \nonumber
                                                      &  + \frac{ 16\pi^3 n^\gamma_{lm}(t)}{ m_a^3 }   \times   [ \frac{ 3 }{  \beta^2} \sum_{l'}  n^a_{l'0}(t)   K^{0}_{l'} - 2\sum_{l'} n^\gamma_{l'0}(t)   N^1_{l'}  ] \}      ~.
\end{flalign}
\begin{flalign}  \tag{\ref{sterile axion number density components equations}  }
& \frac{d n^{as}_{lm}(t) }{dt}        =    \frac{   \Gamma_{a} }{  m_a^2 \beta^2  }   \frac{ 16\pi^3 n^\gamma_{lm}(t)}{ m_a^3 }   \times   2\sum_{l'} n^\gamma_{l'0}(t)   S^1_{l'}     ~.
\end{flalign}

\section{Appendix D: }
Equation \eqref{normal axion number density components equations} for axion number density component $n^a_{00}(t)$ gives
\begin{flalign}  \nonumber  
 \frac{d n^a_{00}(t) }{dt}        =&   - \frac{   \Gamma_{a} }{  m_a^2 \beta^2  }   \{  \sqrt\pi  [  \frac{6 }{ \beta}  \sum_{l'}    K^{0}_{l'}   n^a_{l'0}(t)   \\ \nonumber 
                                                      &  + \frac{ 96\pi^3 }{ m_a^3 \beta^2}  \sum_{l'l''}   n^a_{l'0}(t)  n^\gamma_{l''0}(t)   K^{01}_{l'l''0}   ]   \\ \nonumber
                                                      &  + \frac{ 16\pi^3 n^\gamma_{00}(t)}{ m_a^3 }      [ \frac{ 3 }{  \beta^2} \sum_{l'}  n^a_{l'0}(t)   K^{0}_{l'} - 2\sum_{l'} n^\gamma_{l'0}(t)   N^1_{l'}  ] \}      ~.
\end{flalign}
But the evolution of $n^a_{00}(t)$ can also be obtained by combining equations \eqref{Example-sin^2-axion00-20 relation} and \eqref{Example-sin^2-axion20},
\begin{flalign}  \nonumber  
 \frac{d n^a_{00}(t) }{dt}        =&    \frac{  16 \pi^3 \Gamma_{a} }{  m_a^5 \beta^2  }    n^\gamma_{00}(t)  \times \frac{\sqrt5 n^\gamma_{20}(t)}{n^\gamma_{00}(t)}    \\ \nonumber
                                                           &\times   [ \frac{ 3 }{  \beta^2} \sum_{l'}  n^a_{l'0}(t)   K^{0}_{l'} - 2\sum_{l'} n^\gamma_{l'0}(t)   N^1_{l'}  ]   ~.
\end{flalign}
An algebraic relation for $n^\gamma_{00}(t)$ can be derived from these two differential equations which is
\begin{flalign}  \label{Example-sin^2-axion00-algebraic relation}  
&    \frac{  16\pi^3 \Gamma_{a}   }{  m_a^5 \beta^2  } n^\gamma_{00}(t)  \times [ \frac{\sqrt5 n^\gamma_{20}(t)}{n^\gamma_{00}(t)} +1 ]   \\ \nonumber
                                                           &\times   [ \frac{ 3 }{  \beta^2} \sum_{l'}  n^a_{l'0}(t)   K^{0}_{l'} - 2\sum_{l'} n^\gamma_{l'0}(t)   N^1_{l'}  ]    \\ \nonumber  
=& - \frac{   \Gamma_{a} }{  m_a^2 \beta^2  }     \sqrt\pi  [  \frac{6 }{ \beta}  \sum_{l'}    K^{0}_{l'}   n^a_{l'0}(t)   \\ \nonumber 
                                                &  + \frac{ 96\pi^3 }{ m_a^3 \beta^2}  \sum_{l'l''}   n^a_{l'0}(t)  n^\gamma_{l''0}(t)   K^{01}_{l'l''0}   ]  ~. 
\end{flalign}\\

The photon evolution equation \eqref{photon number density components equations} for component $n^\gamma_{00}(t)$ gives
\begin{flalign}  \nonumber  
 \frac{d n^\gamma_{00}(t) }{dt}        =&    \frac{  2 \Gamma_{a} }{  m_a^2 \beta^2  }   \{  \sqrt\pi  [  \sum_{l'} \frac{6 }{ \beta}   K^{0}_{l'}   n^a_{l'0}(t)   \\ \nonumber 
                                                      &  + \frac{ 96\pi^3 }{ m_a^3 \beta^2}  \sum_{l'l''}   n^a_{l'0}(t)  n^\gamma_{l''0}(t)   K^{01}_{l'l''0}   ]   \\ \nonumber
                                                      &  + \frac{ 16\pi^3 n^\gamma_{00}(t)}{ m_a^3 }   \times   [ \frac{ 3 }{  \beta^2} \sum_{l'}  n^a_{l'0}(t)   K^{0}_{l'}      \\ \nonumber
                                                      &- 2\sum_{l'} n^\gamma_{l'0}(t)   B^1_{l'}  ] \}   - \frac{3c}{2R}n^\gamma_{00}(t)  ~.
\end{flalign}
Substituting relation \eqref{Example-sin^2-axion00-algebraic relation} into this equation, we have
\begin{flalign}  \nonumber  
 \frac{d n^\gamma_{00}(t) }{dt}        =&    \frac{  2 \Gamma_{a} }{  m_a^2 \beta^2  }   \{  -\frac{ 16\pi^3 n^\gamma_{00}(t)}{ m_a^3 }  \times [ \frac{\sqrt5 n^\gamma_{20}(t)}{n^\gamma_{00}(t)} +1 ]   \\ \nonumber
                                                           &\times   [ \frac{ 3 }{  \beta^2} \sum_{l'}  n^a_{l'0}(t)   K^{0}_{l'} - 2\sum_{l'} n^\gamma_{l'0}(t)   N^1_{l'}  ]    \\ \nonumber
                                                      &  + \frac{ 16\pi^3 n^\gamma_{00}(t)}{ m_a^3 }     [ \frac{ 3 }{  \beta^2} \sum_{l'}  n^a_{l'0}(t)   K^{0}_{l'}     \\ \nonumber
                                                      &  - 2\sum_{l'} n^\gamma_{l'0}(t)   B^1_{l'}  ] \}    - \frac{3c}{2R}n^\gamma_{00}(t)  ~.
\end{flalign}
Moving the surface loss term $- \frac{3c}{2R}n^\gamma_{00}(t)$ to the LHS of the equation, combining common factors and noting that $B^1_{l'}=N^1_{l'}+S^1_{l'}$, we have
\begin{flalign}  \nonumber  
& (\frac{d  }{dt}  +  \frac{3c}{2R})         n^\gamma_{00}(t)      \\ \nonumber
 =&    \frac{  2 \Gamma_{a} }{  m_a^2 \beta^2  }   \{  -\frac{ 16\pi^3 n^\gamma_{00}(t)}{ m_a^3 }  \times   \frac{\sqrt5 n^\gamma_{20}(t)}{n^\gamma_{00}(t)}   \\ \nonumber
                                                           &\times   [ \frac{ 3 }{  \beta^2} \sum_{l'}  n^a_{l'0}(t)   K^{0}_{l'} - 2\sum_{l'} n^\gamma_{l'0}(t)   N^1_{l'}  ]    \\ \nonumber
                                                      &  + \frac{ 16\pi^3 n^\gamma_{00}(t)}{ m_a^3 }   \times   [  - 2\sum_{l'} n^\gamma_{l'0}(t)   S^1_{l'}  ] \}  ~. \end{flalign}
Using $N^1_{l'}=B^1_{l'}-S^1_{l'}$ inside the first square bracket leads to
\begin{flalign}  \label{Example-sin^2-1}  
& (\frac{d  }{dt}  +  \frac{3c}{2R}) n^\gamma_{00}(t)  \\ \nonumber
 =&    \frac{  2 \Gamma_{a} }{  m_a^2 \beta^2  }   \{  \frac{ 16\pi^3 [ -\sqrt5 n^\gamma_{20}(t) ]}{ m_a^3 }        \\ \nonumber
                                                           &\times   [ \frac{ 3 }{  \beta^2} \sum_{l'}  n^a_{l'0}(t)   K^{0}_{l'} - 2\sum_{l'} n^\gamma_{l'0}(t)   B^1_{l'}  ]    \\ \nonumber
                                                           &+  \frac{ 16\pi^3 [ -\sqrt5 n^\gamma_{20}(t) ]}{ m_a^3 }  \times  2\sum_{l'} n^\gamma_{l'0}(t)   S^1_{l'}   \\ \nonumber
                                                      &  + \frac{ 16\pi^3 n^\gamma_{00}(t)}{ m_a^3 }   \times   [  - 2\sum_{l'} n^\gamma_{l'0}(t)   S^1_{l'}  ] \}                        ~.     
\end{flalign}\\

The photon evolution equation \eqref{photon number density components equations} for component $n^\gamma_{20}(t)$ gives
\begin{flalign}  \nonumber  
 \frac{d n^\gamma_{20}(t) }{dt}        =&    \frac{  2 \Gamma_{a} }{  m_a^2 \beta^2  }   \times \frac{ 16\pi^3 n^\gamma_{20}(t)}{ m_a^3 }   \times  \\ \nonumber
                                                                      &    [ \frac{ 3 }{  \beta^2} \sum_{l'}  n^a_{l'0}(t)   K^{0}_{l'} - 2\sum_{l'} n^\gamma_{l'0}(t)   B^1_{l'}  ]       - \frac{3c}{2R}n^\gamma_{20}(t)  ~.
\end{flalign}
Move the surface loss term $- \frac{3c}{2R}n^\gamma_{20}(t)$ to the LHS of the equation, 
\begin{flalign}  \label{Example-sin^2-2} 
 (\frac{d  }{dt}  +  \frac{3c}{2R}) n^\gamma_{20}(t)        =&    \frac{  2 \Gamma_{a} }{  m_a^2 \beta^2  }   \times \frac{ 16\pi^3 n^\gamma_{20}(t)}{ m_a^3 }   \times  \\ \nonumber
                                                                      &    [ \frac{ 3 }{  \beta^2} \sum_{l'}  n^a_{l'0}(t)   K^{0}_{l'} - 2\sum_{l'} n^\gamma_{l'0}(t)   B^1_{l'}  ]        ~.
\end{flalign}
Substituting equation \eqref{Example-sin^2-2} into \eqref{Example-sin^2-1} yields
\begin{flalign}  \nonumber  
& (\frac{d  }{dt}  +  \frac{3c}{2R})   \,      n^\gamma_{00}(t)      \\ \nonumber
 =&    (\frac{d  }{dt}  +  \frac{3c}{2R})   \,    [ -\sqrt5 n^\gamma_{20}(t) ]  \\ \nonumber
                                                           &   - \frac{ 32 \pi^3 \Gamma_{a}}{ m_a^5 \beta^2}[ \sqrt5 n^\gamma_{20}(t) + n^\gamma_{00}(t) ]  \times   2\sum_{l'} n^\gamma_{l'0}(t)   S^1_{l'}  ~.  
\end{flalign}
This means that the quantity $n^\gamma_{00}(t) + \sqrt5 n^\gamma_{20}(t)$ evolves based on the following equation,
\begin{flalign}  \nonumber  
&  \frac{d  }{dt}           [ n^\gamma_{00}(t) + \sqrt5 n^\gamma_{20}(t)   ]      \\ \nonumber
                                                           =&   -  [  \frac{ 64\pi^3 \Gamma_{a} }{ m_a^5 \beta^2 }   \sum_{l'} n^\gamma_{l'0}(t)   S^1_{l'}  + \frac{3c}{2R}   ]  [ n^\gamma_{00}(t) + \sqrt5 n^\gamma_{20}(t)   ]  ~.
\end{flalign}

\vspace{0.2cm} 
\begin{acknowledgments}

\end{acknowledgments}

\end{document}